\documentclass[%
 reprint,aip,
superscriptaddress,
 amsmath,amssymb,
]{revtex4-2}
\usepackage{graphicx}
\usepackage{dcolumn}
\usepackage{bm}
\usepackage{soul}
\usepackage[colorlinks]{hyperref}

\usepackage[utf8]{inputenc}
\usepackage[T1]{fontenc}
\usepackage{mathptmx}
\usepackage{etoolbox}
\usepackage{xcolor}
\usepackage{lineno}
\usepackage{amsmath}
\usepackage{amssymb}
\usepackage{mathtools}
\usepackage{cleveref}
\usepackage[english]{babel}
\usepackage{color}

\makeatletter
\def\@email#1#2{%
 \endgroup
 \patchcmd{\titleblock@produce}
  {\frontmatter@RRAPformat}
  {\frontmatter@RRAPformat{\produce@RRAP{*#1\href{mailto:#2}{#2}}}\frontmatter@RRAPformat}
  {}{}
}%

\makeatother
\begin{document}

\preprint{AIP/123-QED}

\title[]{Emergence of collapsed snaking related dark and bright Kerr dissipative solitons with quartic-quadratic dispersion}
\author{Edem Kossi Akakpo}
\affiliation{OPERA-photonics, Universit\'e libre de Bruxelles, 50 Avenue F. D. Roosevelt, CP 194/5, B-1050 Bruxelles, Belgium}
\author{Marc Haelterman}
\affiliation{OPERA-photonics, Universit\'e libre de Bruxelles, 50 Avenue F. D. Roosevelt, CP 194/5, B-1050 Bruxelles, Belgium}
\author{Francois Leo}
\affiliation{OPERA-photonics, Universit\'e libre de Bruxelles, 50 Avenue F. D. Roosevelt, CP 194/5, B-1050 Bruxelles, Belgium}
\author{Pedro Parra-Rivas}%
 \email{pedro.parra-rivas@uinroma1.it}
\affiliation{Dipartimento di Ingegneria dell’Informazione$,$ Elettronica e Telecomunicazioni$,$
Sapienza Universit\'a di Roma$,$ via Eudossiana 18$,$ 00184 Rome}%
\affiliation{OPERA-photonics, Universit\'e libre de Bruxelles, 50 Avenue F. D. Roosevelt, CP 194/5, B-1050 Bruxelles, Belgium}

\begin{abstract}
We theoretically investigate the dynamics, bifurcation structure and stability of dark localized states emerging in Kerr cavities in the presence of second- and fourth-order dispersion. These states form through the locking of uniform wave fronts, or domain walls, connecting two coexisting stable uniform states. They undergo a generic bifurcation structure known as collapsed homoclinic snaking. We characterize the robustness of these states by computing their stability and bifurcation structure as a function of the main control parameter of the system. Furthermore, we show that by increasing the dispersion of fourth order, bright localized states can be also stabilized.
\end{abstract}
\maketitle




\section{Introduction}

The generation of dissipative solitons, also referred to as localized states (LSs), in externally driven dispersive Kerr optical resonators hinges in a double balance condition between four different factors which compensate pairwise: Kerr nonlinearity counteracts chromatic dispersion (similarly to single pass conservative systems), while energy dissipation is compensated through external driving \cite{akhmediev_dissipative_2005}. The dynamics of the cavity can be basically controlled by tuning the external energy source. However, in most of the applications, it is extremely important the careful management of dispersion, which has become a fundamental engineering problem. In microresonators, for example, the dispersion can be engineered through the resonator geometry \cite{kippenberg_dissipative_2018}.  Recently, a new accurate control of dispersion, based on an intracavity pulse shaper, 
was proposed by Runge {\it et al.} in soliton lasers, opening new avenues regarding LSs formation and control \cite{runge_pure-quartic_2020}. 

Generally, the formation of dissipative solitons in Kerr resonators is based on the concept of bistability and front locking: when two different stable states coexist for the same range of parameters, front waves may form, interact and lock, leading to the formation of a plethora of LSs of different extensions \cite{thual_localized_1988,coullet_localized_2002}. This mechanism has been useful to explain the formation of LSs based on second-order dispersion (SOD) \cite{parra-rivas_dark_2016,parra-rivas_frequency_2019,parra-rivas_localized_2019,arabi_localized_2020,parra-rivas_dissipative_2022}, but also to understand the implications that higher-order dispersion effect may have on the LS dynamics and stability \cite{parra-rivas_coexistence_2017}. One of the first studies in this topic, showed that fourth-order dispersion (FOD) was able to stabilize dark localized pattern in a regime where otherwise such states were absent \cite{tlidi_high-order_2010}. Later, different studies have tackled the effect of third-order dispersion (TOD) on different bistable configurations, showing that, in general, new type of bright solitons can be formed \cite{parra-rivas_third-order_2014,parra-rivas_coexistence_2017}.

In the last few years, dissipative solitons have been also studied in Kerr cavities driven at the pure FOD point \cite{bao_high-order_2017,taheri_quartic_2019}, and their bifurcation structure and stability analyzed in the anomalous and normal regimes \cite{parra-rivas_quartic_2022}. In the first case, single and multi-peak LSs persist and are stable over a wider parameter region than those in the pure SOD case. Moreover, in the second scenario, pure FOD is able to stabilize bright LSs.  Thus, in general, higher-dispersion effects seem to have a positive impact on soliton formation and stability.  

\begin{figure}[!t]
\centering
\includegraphics[scale = 1]{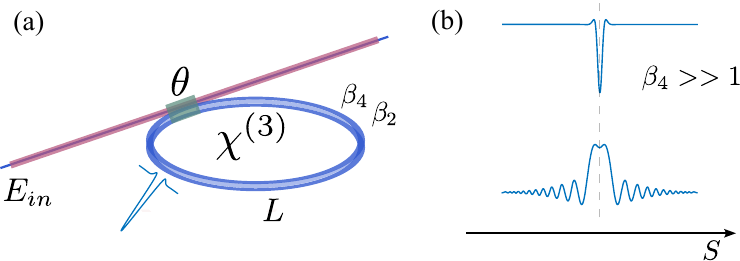}
\caption{(a) Sketch of a externally driven Kerr resonator of length $L$. $\theta$ represents the transmission coefficient of the coupler between the $E_{in}$ and the cavity. $\beta_2$ and $\beta_4$ are the second- and fourth-order dispersion coefficients (b) Coexistence of bright and dark LSs for large $\beta_4$.}
\label{fig0}
\end{figure}
In this work, we analyze a externally driven Kerr resonator [see sketch in Fig.~\ref{fig0}(a)] in a operation regime not tackled previously, were both normal SOD and FOD effects contribute to the cavity dynamics. We show that in this regime, the locking of fronts connecting two uniform bistable states leads to the formation of collapsed snaking related dark LSs \cite{knobloch_homoclinic_2005}. This configuration shows morphological similarities with the normal SOD scenario \cite{parra-rivas_dark_2016}, despite the different dynamical instabilities thresholds and the LSs existence domains. Increasing the FOD strength, we find that different types of bright LSs are stabilized and coexist with dark ones [see Fig.~\ref{fig0}(b)], in a fashion similar to the one described in the context of TOD \cite{parra-rivas_coexistence_2017}. In what follows, we unveil the bifurcation structure organization associated with such coexistence, showing the main dynamical regimes of the system.

This paper is organized as follows. In Sec.~\ref{sec:1} we introduce the model and give some preliminary results. We present the time-independent problem (Sec.~\ref{sec:1.1}), show the homogenous steady state solution (Sec.~\ref{sec:1.2}), and compute its linear stability (Sec.~\ref{sec:1.3}). Section~\ref{sec:2} focuses on the bifurcation structure and stability of dark LSs and in Sec.~\ref{sec:3} we study the spatio-temporal dynamics of some of the LSs. The effect that the variation of FOD may have on the dynamics and stability of the dark LSs and the stabilization of bright LSs is analyzed in Sec.~\ref{sec:4}. Finally, we discuss the results and draw some conclusions in  Sec.~\ref{sec:5}.

\section{The Lugiato-Lefever model with quartic dispersion}\label{sec:1}

In the mean-field approximation, passive Kerr cavities can be described by the Lugiato-Lefever (LL) equation \cite{haelterman_dissipative_1992}. Considering chromatic dispersion up to fourth-order, and neglecting the contribution of TOD, the normalized LL equation reads
\begin{equation}\label{LLE}
\partial_t A = -(1 + i \Delta) A - i d_2 \partial_x^2 A + i d_4 \partial_x^4 A + i |A|^2 A + S,
\end{equation}
where $A$ is the complex field amplitude, $t$ represents the time coordinate, and $x$ the fast time in fiber cavities or angular variable in microresonators. The losses are normalized to 1, $\Delta$ is the phase detuning from the closest cavity resonance and $S$ in the driving field amplitude. With this normalization,  $d_2={\rm sign}(\beta_2)=\pm 1$ and $d_4\equiv\beta_4 \alpha/(6L|\beta_2|^2)$, where $\alpha$ represent the losses, $L$ is the cavity length, and $\beta_2$ and $\beta_4$ represent the SOD and FOD coefficients, respectively. In this work we focus on the regime defined by $d_2=1$   and $d_4>0$.

\subsection{The time-independent problem: Spatial dynamics}\label{sec:1.1}


LSs and any other type of time-independent states ($\partial_t A = 0$) satisfy 
the complex ordinary differential equation
\begin{equation}\label{sta}
i d_4 \partial_x^4 A = i d_2 \partial_x^2 A +  (1 + i \Delta) A - i |A|^2 A - S.
\end{equation}
Considering $A = U + i V$, Eq.~(\ref{sta}) can be recast into the 8D spatial dynamical system  
\begin{equation}\label{DS}
\frac{d {\bf Y}}{dx} = {\bf F}({\bf Y}, d_2, d_4, \Delta, S),
\end{equation}
where the new variables read $${\bf Y}=(Y_1,Y_2,\cdots,Y_8)=(U,V,\partial_x U, \partial_x V,\partial^2_x U, \partial^2_x V,\partial^3_x U, \partial^3_x V),$$
and the vector field {\bf F} is defined as
\begin{equation*}
\left\{
\begin{split}
F_m &= Y_{m+2}, \hspace{2cm}{m=1,\cdots, 6} \\
F_7 &= \frac{1}{d_4}[\Delta  Y_1  - (Y_1^2 + Y_2^2) Y_1 + Y_2 + d_2 Y_5]  \\
F_8 &= -\frac{1}{d_4} [Y_1 + \Delta Y_2 - (Y_1^2 + Y_2^2) Y_2 + d_2Y_6 + S]  \\
\end{split}
\right.
\end{equation*}

By using the dynamical system (\ref{DS}), we can apply well known results of dynamical systems and bifurcation theory to study our problem. Furthermore, this approach allows us to apply a correspondence between the time-independent solutions of the system [i.e., solutions of Eq.~(\ref{sta})] and different solutions of Eq.~(\ref{DS}). In this context, a uniform front corresponds to a {\it heteroclinic orbit}, a LS is a {\it homoclinic orbit}, a spatially periodic pattern a {\it limit cycle}, and the homogeneous state of the system a {\it fixed point} \cite{parra-rivas_origin_2021}.

Equation~\ref{DS} will be used for computing LSs through numerical path-continuation algorithms based on a predictor-corrector method \cite{doedel_numerical_1991,doedel_numerical_1991-1} using the free software package AUTO-07p \cite{Doedel2009}. With this procedure, we are able to compute not only the stable but also the unstable state solutions, unveiling their connection. 
The stability is determined after continuation by solving the eigenvalue problem
\begin{equation}
    \mathcal{L}\psi=\sigma\psi,
\end{equation}
where $\mathcal{L}$ is the linear operator associated with the right hand side of Eq.~(\ref{LLE}) evaluated a given steady state, and $\psi$ is the eigenmode corresponding to the eigenvalue $\sigma$. The time-independent state is stable if Re$[\sigma]<0$, and unstable otherwise. If by varying a control parameter of the system, let's say $p$, this transition occurs at the value $p=p_c$ (i.e., Re$[\sigma(p_c)]=0$), we say that a local bifurcation takes place at $p_c$ \cite{wiggins_introduction_2003,scroggie_pattern_1994}.

\begin{figure}[!t]
\centering
\includegraphics[scale = 1]{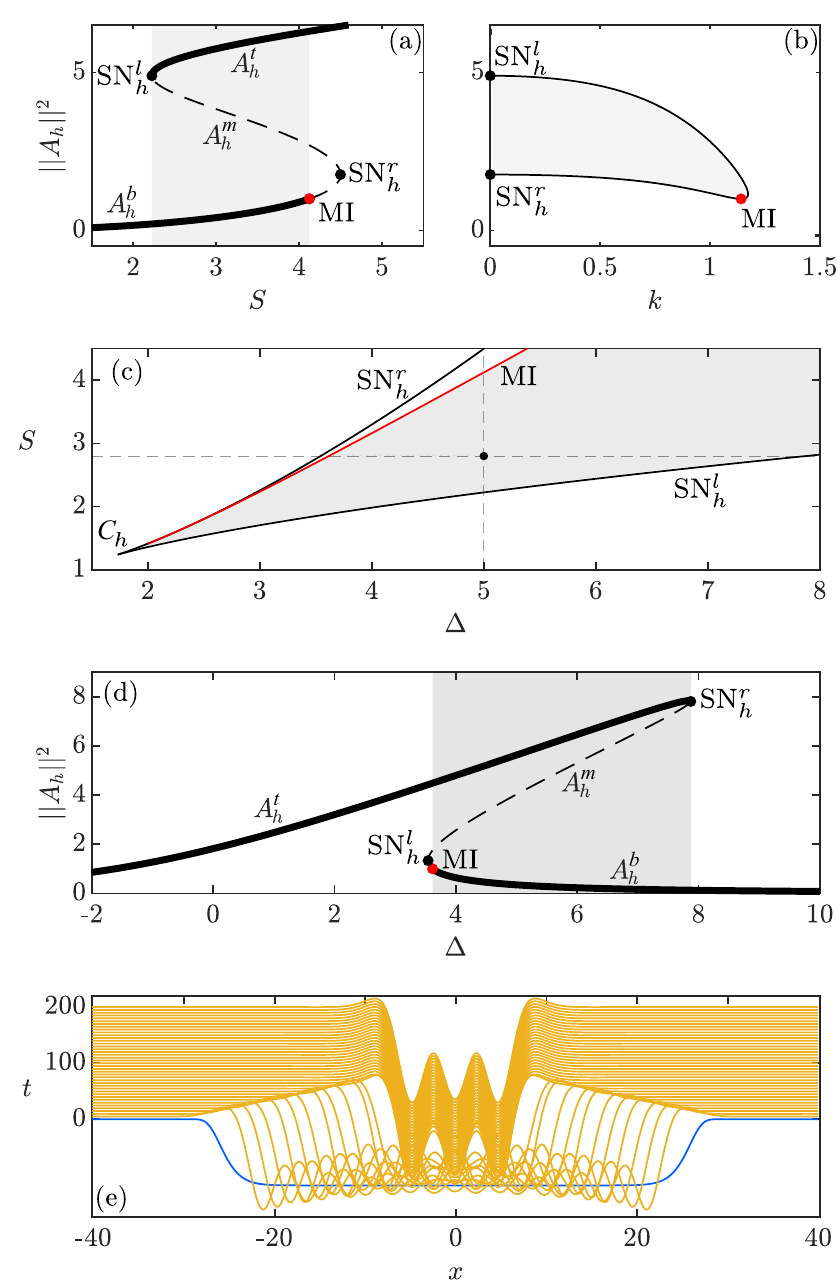}
\caption{(a) Homogeneous steady state in the bistable regime for $\Delta=5$. (b) Marginal instability curve corresponding to (a). Panel (c) shows the $(\Delta,S)$-parameter space. The vertical line corresponds to the situation shown in (a). The horizontal line to the nonlinear cavity resonance shown in (d) for $S=2.8$. (e) Shows the formation of a dark LS for $\Delta=5$ and $S=2.8$. The blue curve in (e) represents the initial condition. }
\label{fig1}
\end{figure}

\subsection{Homogeneous steady state}\label{sec:1.2}
The simplest time-independent solution is the uniform or homogeneous steady state (HSS) solution $A_h$ (i.e., $\partial_x A=0$), which satisfies the algebraic equation 
\begin{equation}
      (1 + i \Delta) A_h - i |A_h|^2 A_h - S=0.
\end{equation}
This equation can be rewritten in the form 
\begin{equation}\label{Sshape}
S^2 = I_h^3 - 2\Delta I_h^2 + (1 + \Delta^2)I_h,
\end{equation}
with $I_h\equiv|A_h|^2$. For a fixed value of $\Delta$, this expression defines a nonlinear dependence between the intracavity intensity $I_h$ and the pump $S$. An example of such dependence is depicted in Fig.~\ref{fig1}(a) for $\Delta=5$.

For this value of $\Delta$, the system shows multistability, i.e., for the same value of $S$ three HSSs, namely $A_h^b$, $A_h^m$, and $A_h^t$, coexist. These states are connected at two folds, or turning points, located at the positions 
\begin{equation}\label{fold_points}
    I_h^{l,r}\equiv\frac{1}{3}\left(2\Delta\pm\sqrt{\Delta^2-3}\right).
\end{equation}
As a function of $\Delta$, these folds points define the two solid lines plotted in the $(\Delta,S)$-parameter diagram shown in Fig.~\ref{fig1}(c). Decreasing $\Delta$, eventually, these two folds meet and disappear in a cusp bifurcation $C_h$ occurring exactly at $\Delta=\sqrt{3}$. The vertical dashed line shown in  Fig.~\ref{fig1}(c) corresponds to the diagram plotted in Fig.~\ref{fig1}(a).

For a fixed value of $S$, Eq.~(\ref{Sshape}) is solved by the expression  
\begin{equation}
    \Delta=I_h\pm\sqrt{(S/I_h)^2-1}.
\end{equation}
An example of these solution branches is plotted in Fig.~\ref{fig1}(d) for  $S=2.8$, and represents the nonlinear resonance of the cavity.

\subsection{Linear stability analysis of the uniform state}\label{sec:1.3}
The linear stability of HSS against perturbations $\sim e^{\sigma t}\psi_k$ can be determined analitically. In this case, eigenmodes read $\psi_k=e^{ikx}+{\rm c.c.}$, while the eigenvalues depend on the wavenumber $k$ through the dispersion relation 
\begin{align}
\sigma(k) = -1 \pm \sqrt{-K(k)^2 - 2 (\Delta - 2I_h)K(k) - C}, && 
\end{align}
with 
\begin{subequations}
    \begin{equation}
        K(k) = -d_2 k^2 - d_4 k^4,
    \end{equation} and
    \begin{equation}
        C = (\Delta - I_h)^2  - 2I_h(\Delta - I_h)
    \end{equation}
\end{subequations}
For non-uniform perturbations $(k\neq0)$, the transition stable/unstable takes place at the Turing or modulational instability (MI), which satisfies simultaneously the conditions $\partial_k\sigma(k)=0$ and $\sigma(k) = 0$ for a critical wavenumber $k=k_c$. From this condition, MI occurs at $I_h=1$, and the growing perturbation at this point has the wavenumber


\begin{equation}
k_c = \pm \sqrt{\frac{-d_2 \pm \sqrt{d_2^2 - 4d_4(2  - \Delta)}}{2 d_4}}, 
\end{equation}
with $d_2=1$ for our regime.


The condition $\sigma(k)=0$ leads also to the marginal instability curve 
\begin{equation}
I_h = \frac{1}{3} \left[ \sqrt{ K^{2} + 2 K \Delta + {{\Delta }^{2}} -3} \pm 2( \Delta + K) \right], 
\end{equation}
which separates stable from unstable regions. 

Figure~\ref{fig1}(b) shows the marginal instability curve corresponding to the the HSS shown in Fig.~\ref{fig1}(a). The minimum of this curve corresponds to the MI, and the area inside it to the unstable HSSs. In correspondence, stable (unstable) HSSs are marked using solid (dashed) lines in Fig.~\ref{fig1}(a). At $k=0$, two homogeneous instabilities [i.e., saddle-node (SN) bifurcations] occur at the fold positions [see Eq.~(\ref{fold_points})]. In what follows we label these points SN$_h^{l,r}$.

\begin{figure}[!t]
\centering
\includegraphics[scale = 1]{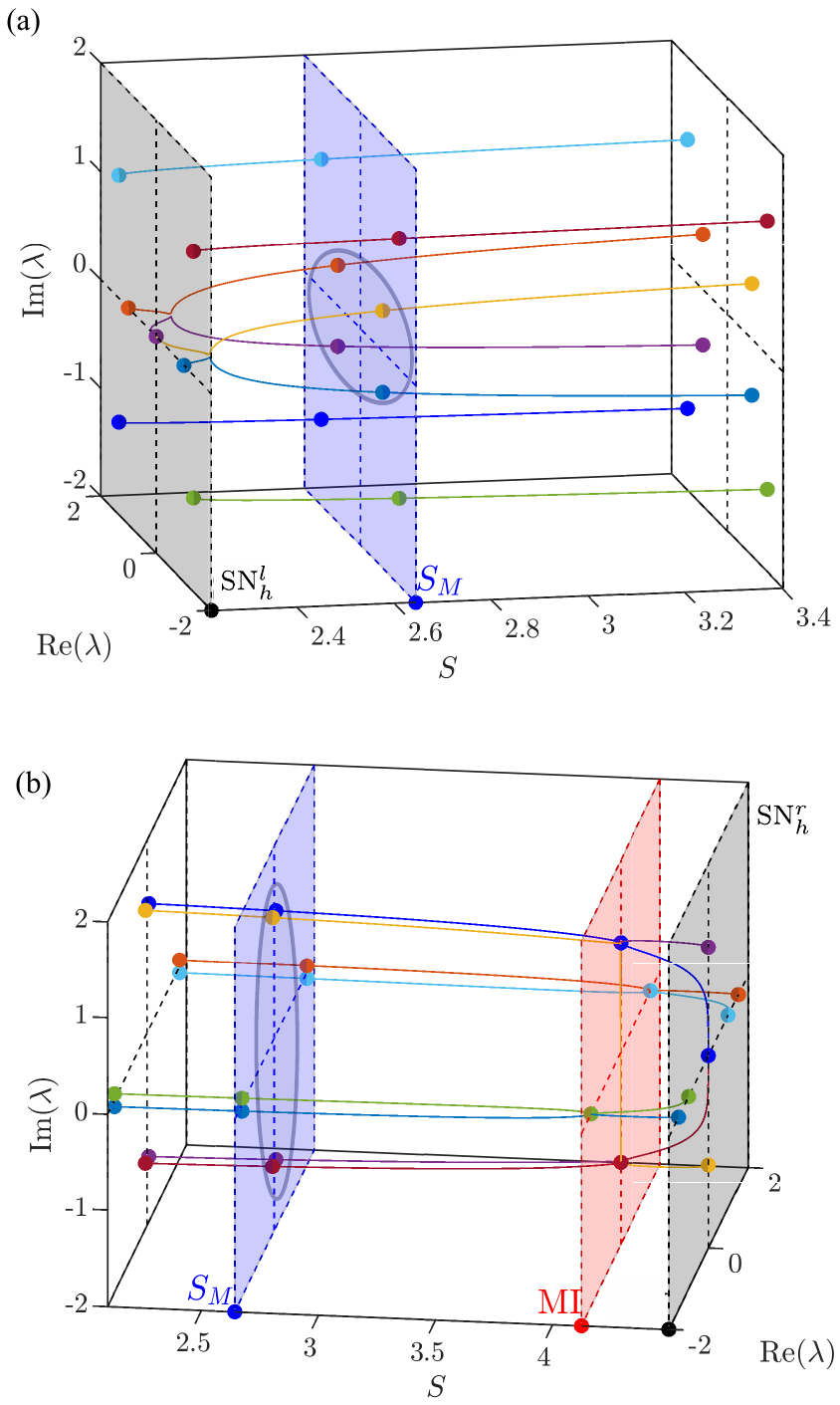}
\caption{(a) Modification of the spatial eigenvalues associated with $A_h^t$ as a function of $S$ for $\Delta=5$. The gray plane corresponds SN$_h^l$, and the blue one to $S_M$. In this plane, the most relevant eigenvalues are surrounded by a gray oval. (b) Same than in (a) but for $A_h^b$.}
\label{eigen_1}
\end{figure}


\section{Bifurcation analysis for dark localized states}\label{sec:2}

\subsection{Bistability and plane-front locking}\label{sec:2.1}
In the bistability region shown in Fig.~\ref{fig1}, fronts connecting $A_h^t$ and $A_h^b$ forwards and backwards can form. These fronts drift at a constant speed which depends on the parameters of the system. The speed increases (decreases) as the system parameters approach (separate) from the Maxwell point of the system, where it cancels out. Around this zero-speed point, fronts can lock if oscillatory tails exist in the front profiles, leading to the formation of LSs of different widths \cite{coullet_nature_1987,coullet_localized_2002}. Asymptotically, these tails can be described through the expression $A(x)-A_h\sim e^{\lambda x}$, where $\lambda$ is the spatial eigenvalue of the system evaluated at $A_h$. The spatial eigenvalues can be computed through the Jacobian associated with Eq.~(\ref{DS}), or equivalently, by solving the equation $\sigma(-i\lambda)=0$, namely
\begin{multline*}
d_4^2 \lambda^8- 2 d_2 d_4 \lambda^6 + (4 I_h d_4 - 2 \Delta d_4  + d_2^2) \lambda^4\\- (4 I_h d_2 + 2 \Delta d_2) \lambda^2 + (\Delta^2 - 4 I_h \Delta + 3 I_h^2 + 1) = 0.
\end{multline*}
In our regime ($d_2 = 1$ and $d_4 = 1$), the previous equation has 8 solutions that read 
\begin{equation}
\lambda= \pm\frac{1}{\sqrt{2}} \sqrt{-1 \pm \sqrt{4 \Delta -8 {I_h}+1 \pm 4 \sqrt{{{{I_h}}^{2}}-1}}} 
\end{equation}
The dynamically relevant eigenvalues are those related with the slow dynamics of the system (i.e., those with smallest $|{\rm Re}[\lambda]|$). Oscillatory tails appear if the dominant eigenvalues are complex-conjugate. When these eigenvalues are all real numbers, the tails are monotonic. The transition between these two configurations depends on the parameters of the system (for this case $\Delta$ and $S$). The modification of the spatial eigenvalues associated with $A_h^t$, and $A_h^b$ as a function of $S$ are depicted in Figs.~\ref{eigen_1}(a) and (b), respectively, for $\Delta=5$.

Let's take a look to the eigenvalues associated with $A_h^{b,t}$ around $S_M$ (see blue planes in Fig.~\ref{eigen_1}). For $A_h^b$, the dominant eigenvalues around $S_M$ [see oval gray shape in Fig.~\ref{eigen_1}(b)] have a large imaginary part and a very small real part, what means that the oscillatory tails will have small wavelength and weak decay. This leads to well defined oscillatory tails around $A_h^b$. For $A_h^t$ [see Figs.~\ref{eigen_1}(a)], although the dominant eigenvalues are also complex conjugate, the imaginary and real parts are respectively smaller and larger than in Figs.~\ref{eigen_1}(b), which leads to oscillatory tails with a very strong decay and very large wavelength: effectively a quasi monotonic tail. Therefore, front locking will be favored around $A_h^b$, in contrast with $A_h^t$.
\begin{figure}[!t]
\centering
\includegraphics[scale = 1]{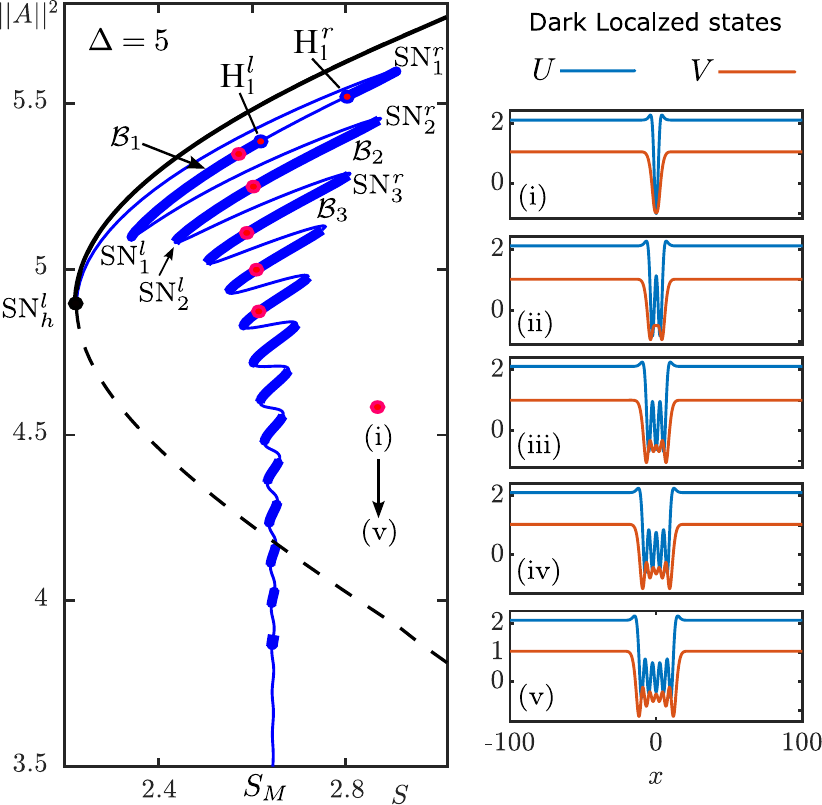}
\caption{Collapsed homoclinic snaking for $\Delta=5$. Solid thick (thin) lines correspond to stable (unstable) states. Some examples of dark LSs are shown on the right panels (i)-(v). $\mathcal{B}_i$ label the solution branches in-between SN$_i^{l,r}$, and $S_M$ is the uniform Maxwell point of the system.}
\label{dia1}
\end{figure}
\begin{figure*}[!t]
\centering
\includegraphics[scale = 0.95]{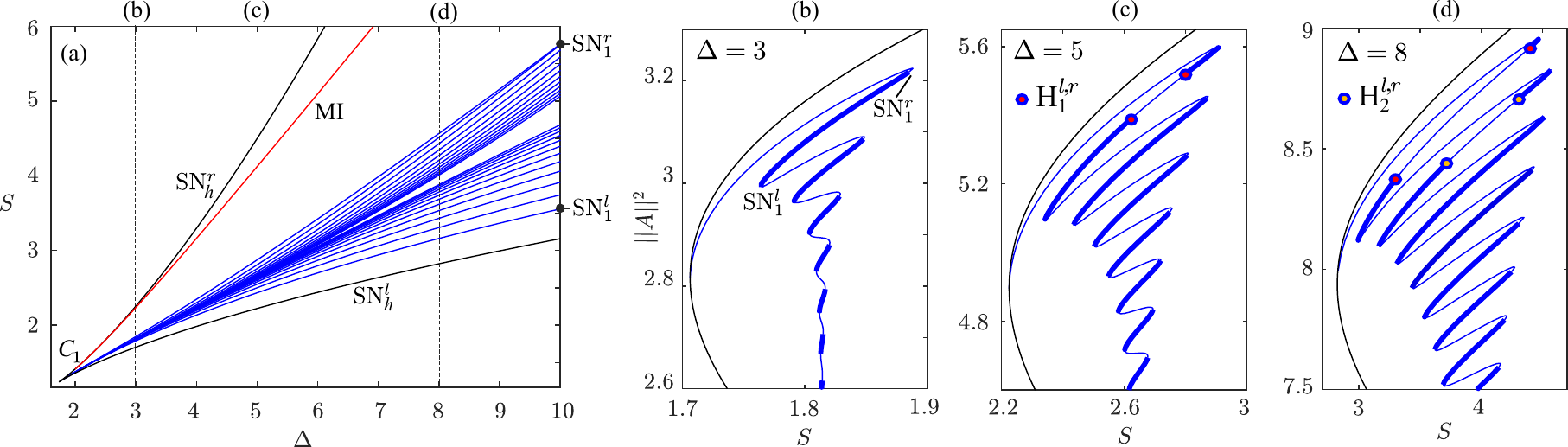}
\caption{(a) $(\Delta,S)$-parameter space for $\beta_4=1$. This diagram shows the main bifurcations of the system: the MI, saddle-node bifurcations of the homogeneous state SN$_h^{l,r}$, and saddle-node bifurcations associated with the collapsed homoclinic snaking SN$_i^{l,r}$. Panels (b)-(d) show the collapsed snaking for $\Delta=3, 5$ and $8$ respectively, corresponding to the vertical dashed lines plotted in (a). Solid thick and thin lines represent stable and unstable LS solutions. The labels H$_i^{l,r}$ mark the position of the Hopf bifurcations leading to breathing behavior. }
\label{Delta_S_diagram}
\end{figure*}

An example of front locking is shown in Fig.~\ref{fig1}(e). This temporal evolution has been computed through a direct numerical simulation of Eq.~(\ref{LLE}) starting from a super-Gaussian profile subtracted to $A_h^t$ close to $S_M$ [shown in blue in Fig.~\ref{fig1}(e)]. Initially, two fronts with opposite polarity form and approach one-another as time passes. Eventually, they lock at a fixed separation $D$, yielding a dark LS with two central dips.  
The time evolution of that separation can be described by the equation \cite{parra-rivas_origin_2021}
\begin{equation}\label{Interaction}
    \partial_t D=\varrho e^{{\rm Re}[\lambda]D}{\rm cos}({\rm Im}[\lambda]D)+\eta,
\end{equation}
where $\varrho$ depends on the parameters and $\eta\sim S-S_M$. Although the front interaction appearing here is generic, and well described by Eq.~(\ref{Interaction}), its derivation from Eq.~(\ref{LLE}) is not realizable, and we introduce it here for explaning the front locking mechanism.   The fixed points $D_e$ of this system ($\partial_t D_e=0$) correspond to the locking of fronts, and hence, to the formation of LSs of width $D_e$. Thus, for the same value of $\eta$, LSs of different widths may coexist. As we will see in the coming section, LSs formed through front locking organize in a particular bifurcation structure which depends directly on the interaction law (\ref{Interaction}).

\subsection{Bifurcation structure: Collapsed homoclinic snaking}\label{sec:2.2}
To fully understand the formation of these states we perform a bifurcation analysis based on the path-continuation techniques described in Sec.~\ref{sec:1.1}. The output of these computations leads the bifurcation diagram shown in Fig.~\ref{dia1} where the energy of $A$, i.e., the $L_2$-norm $$||A||^2\equiv L^{-1}\int_{-L/2}^{L/2}|A(x)|^2 dx$$ is plotted as a function of $S$ for $\Delta=5$. This diagram is known as {\it collapsed homoclinic snaking} and is generic of systems where LSs emerge through uniform-front locking. Indeed, the damped oscillatory shape of the bifurcation curve around $S_M$ is a direct consequence of the front interaction and locking described by Eq.~(\ref{Interaction}) (see Ref.~\cite{parra-rivas_origin_2021} for a general description). The collapsed snaking curve emanates from SN$_h^l$. Close to this bifurcation, LSs have small amplitude and are unstable. These small amplitude states can be computed through multi-scale perturbation theory as done in the standard LL equation \cite{parra-rivas_dark_2016,parra-rivas_origin_2021}. Following the diagram downwards, the dark LSs undergo a sequence of saddle-node bifurcations appearing in pairs SN$_i^{r,l}$, where the sub-index $i$ represents the number of dips appearing in each state. In these bifurcations, LSs gain and loss stability, and at each SN$_i^l$ an extra dip is nucleated in the structure at $x=0$. In this way, the width of the LSs increases while decreasing $||A||^2$. In what follows, we mark the solution branches connecting SN$_i^{l,r}$ as $\mathcal{B}_i$ (see Fig.~\ref{dia1}).

\subsection{Persistence in the $(\Delta,S)$-parameter space}\label{sec:2.3}

\begin{figure*}[!t]
\centering
\includegraphics[scale = 1]{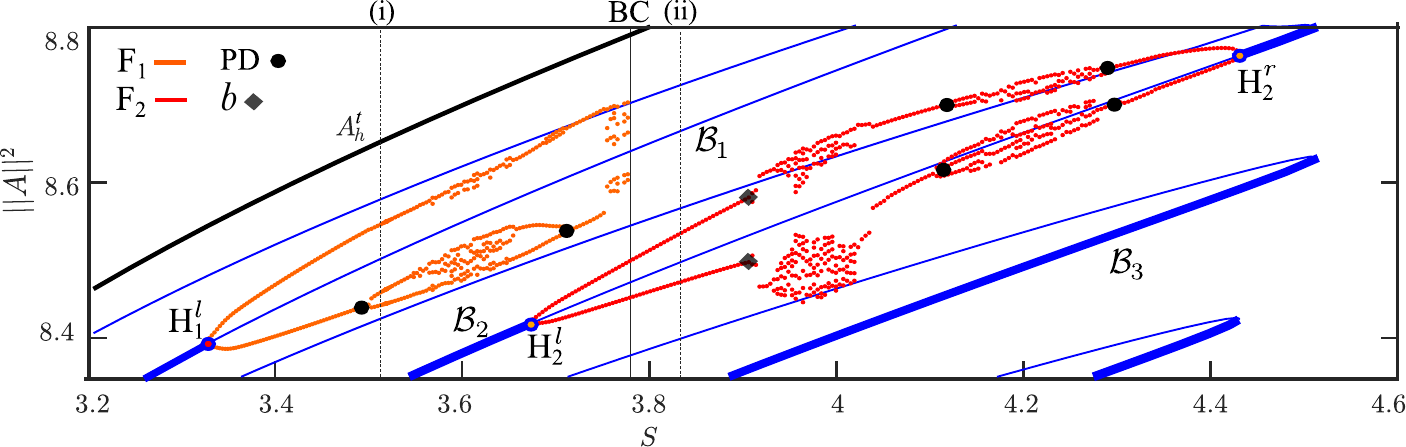}
\includegraphics[scale = 0.9]{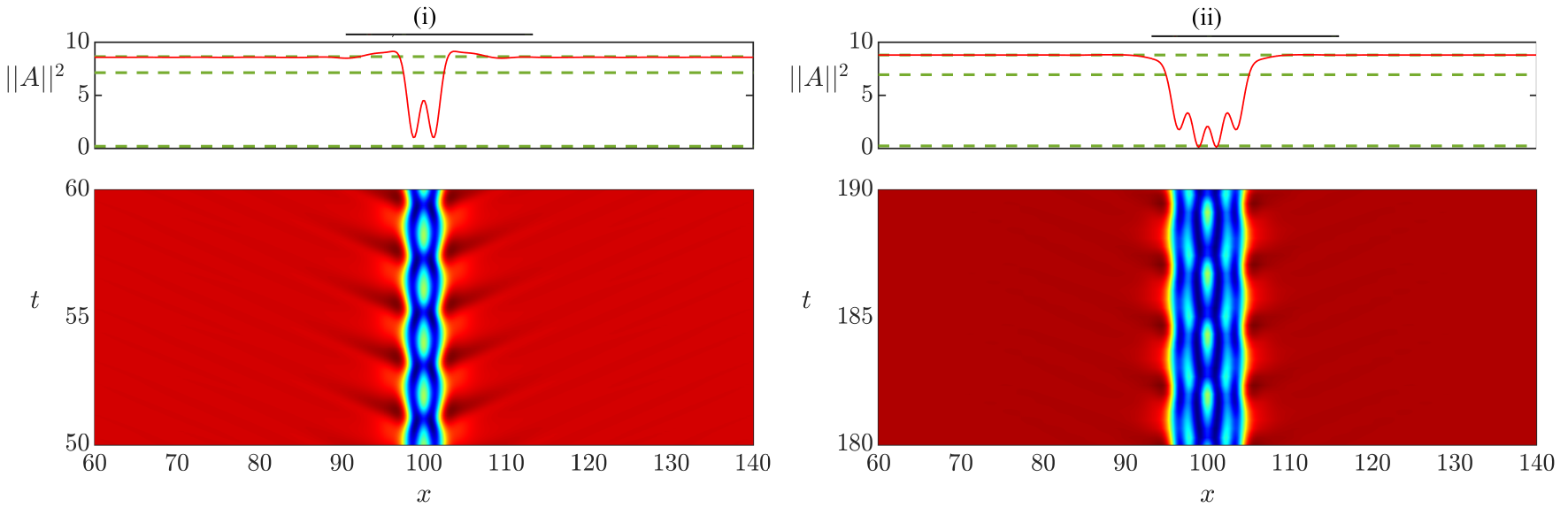}
\caption{Top part of the collapsed snaking diagram for $\Delta=8$ where we show all the local maxima and minima of the breathers. We use orange dots for the breathers emerging from $\mathcal{B}_1$ at H$_1^{l,r}$ (i.e., F$_1$), and red dots for those bifurcating from H$_2^{l,r}$ along $\mathcal{B}_2$ (i.e., F$_2$). PD corresponds to a period-doubling bifurcation, and $b$ marks the emergence of beating behavior. Panels (i) and (ii) show two examples of breathers corresponding to  F$_1$ and F$_2$, respectively. }
\label{breathers1}
\end{figure*}

Figure~\ref{Delta_S_diagram}  shows the modification of the  SN$_i^{l,r}$ bifurcations (see blue lines) in the $(\Delta,S)$-parameter space, together to SN$_h^{l,r}$ and MI. With decreasing $\Delta$, the pairs SN$_i^{l,r}$ approach each other and eventually they meet in a sequence of cusp bifurcations $C_i$ (here we only show $C_1$). Those with larger $i$, corresponding to SNs down in the collapsed snaking shown in Fig.~\ref{dia1}, are the one disappearing first, when decreasing $\Delta$. 
This phase diagram is common in systems sharing similar HSS linear stability, such as in passive Kerr cavities with second-order chromatic dispersion \cite{parra-rivas_dark_2016} and in dispersive cavity enhanced second-harmonic generation \cite{parra-rivas_dark_2021}.

The modification of the collapsed snaking diagram along the $(\Delta,S)$-parameter space is illustrated in Fig.~\ref{Delta_S_diagram}(b)-(d) for $\Delta=3$, $5$ and $8$, respectively [see vertical dashed lines in Fig.~\ref{Delta_S_diagram}(a)]. For $\Delta=3$, all the $\mathcal{B}_i$ branches are stable, and most of them correspond to narrow states. Increasing $\Delta$, cusp $C_i$ with larger $i$ appear, and  wider states emerge. Moreover, $\mathcal{B}_i$ become wider due to the separation of their corresponding SN$^{l,r}_i$.

The linear stability analysis along these diagrams shows that, eventually, $\mathcal{B}_{1}$ undergoes a pair of Hopf bifurcaitons H$_1^{l,r}$, where the single-dip LS becomes unstable in favor of localized oscillations (i.e., breathers). This is the situation depicted in Fig.~\ref{Delta_S_diagram}(c) for $\Delta=5$. We will explore the spatiotemporal dynamics of these states in  Sec.~\ref{sec:3}.

Further increasing $\Delta$, $\mathcal{B}_{2}$ also destabilizes through the Hopf bifurcations H$_2^{l,r}$ [see Fig.~\ref{Delta_S_diagram}(d) for $\Delta=8$], leading to the appearance of new breather states.
A formal comparison with the 
standard LL equation in the normal chromatic dispersion regime (see Ref.~\cite{parra-rivas_dark_2016}) shows that the destabilization of the  $\mathcal{B}_{i}$ branches occurs for larger values of $\Delta$.

\section{Spatio-temporal dynamics}\label{sec:3}

\begin{figure*}[!t]
\centering
\includegraphics[scale = 1]{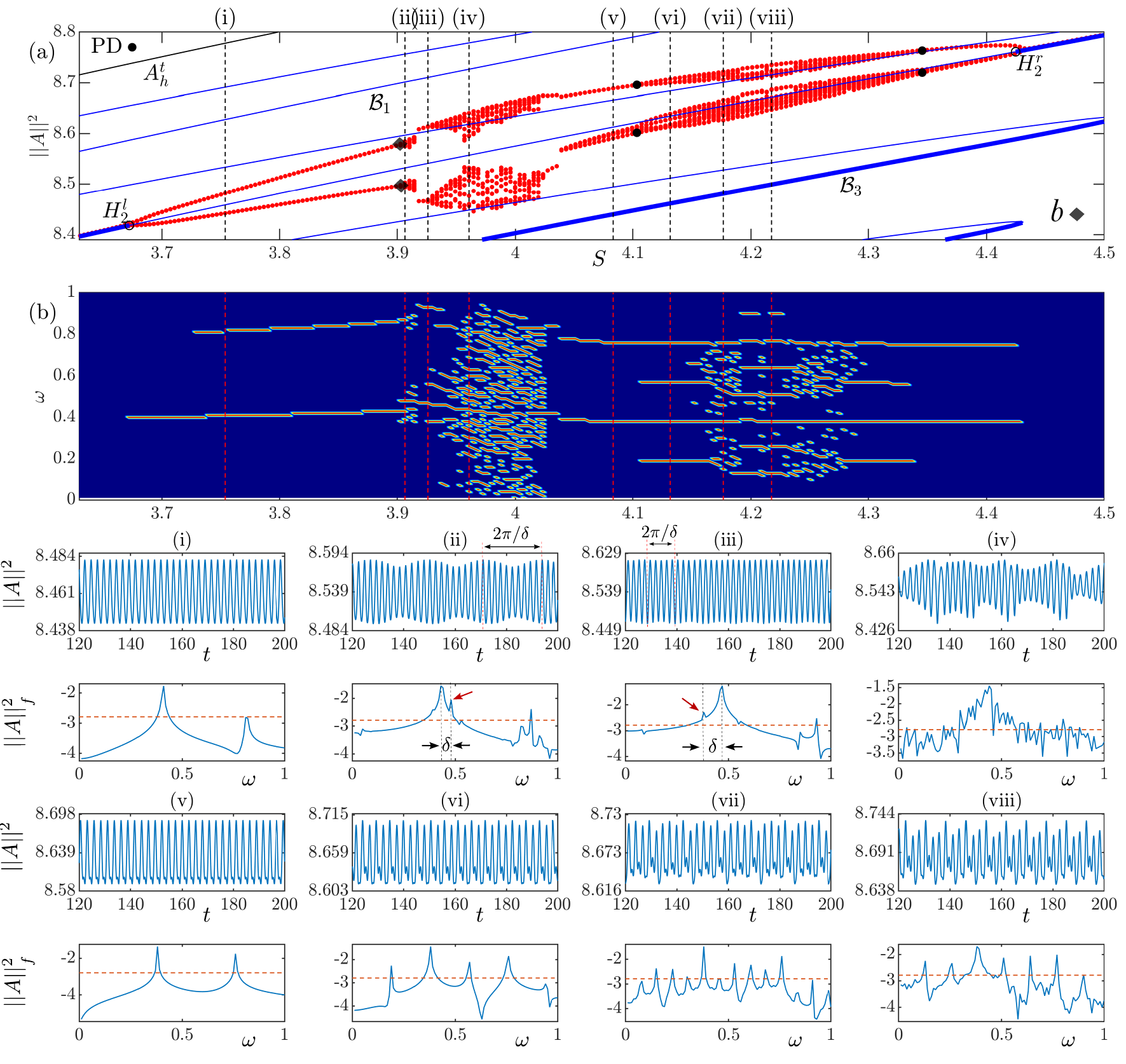}
\caption{(a) Close-up view of the Fiegenbaum-like diagram F$_2$ between H$_2^{l,r}$  (see Fig.~\ref{breathers1}). (b) Modification of the frequency spectrum of the oscillatory states along the diagram shown in (a). To plot this spectrum we have filtered out the frequencies below the red dashed line depicted in panels (i)-(vi). These panels show the time series and frequency spectra of the different oscillatory states [see vertical dashed lines in (a) and (b)]. }
\label{breather2}
\end{figure*}


Figure~\ref{breathers1} shows a portion of the bifurcaiton diagram depicted in Fig.~\ref{Delta_S_diagram}(d) for $\Delta=8$, where the maxima and minima of the oscillatory variation of the breathers' norm emerging from H$_1^{l,r}$ are plotted using orange circles, whereas those emanating from H$_2^{l,r}$ are shown using red circles. In what follows, we refer to this set of breather branches as Feigenbaum-like diagrams (F) \cite{ott_chaos_2002}: F$_1$ (orange dots) is connected to $\mathcal{B}_1$, while F$_2$ forms along $\mathcal{B}_2$ (red dots). These diagrams, and therefore the breathers, bifurcate supercritically from every H$_i^{l,r}$.

Let us first focus on F$_1$. An example of the breather belonging to this set is depicted in Fig.~\ref{breathers1}(i) for $S=3.52$. This state emerges with a small amplitude form H$_1^l$ and with a single oscillatory period. The breather undergoes a sequence of period-doubling (PD) bifurcations yielding more complex dynamics, that we do not show here. The PD cascade is observable from the F$_1$ diagram, although we only mark the first one at each side. Near $\Delta\approx3.6$, the system evolves to temporal chaos, and this PD sequence is inverted. Eventually the chaotic state dies, possibly in a boundary crisis of the attractor (BC) \cite{ott_chaos_2002}, and the system ends up in the closest basin of attraction: the HSS $A_h^t$. This situation share similarities with the one reported in Kerr cavities when only second-order dispersion is considered \cite{parra-rivas_dark_2016}. 

The F$_2$ diagram in Fig.~\ref{breathers1} shows more complexity than F$_1$, and furthermore, does not encounter any crisis or instability destroying the oscillatory states. An example of a single-period breather bifurcating from H$_2^l$ is depicted in Fig.~\ref{breathers1}(ii) for $S=3.82$. Here, F$_2$ shows the transition between different dynamical regimes including beating phenomena ($b$) between different frequencies and PD cascades \cite{ott_chaos_2002}.
\begin{figure*}[!t]
\centering
\includegraphics[scale =1]{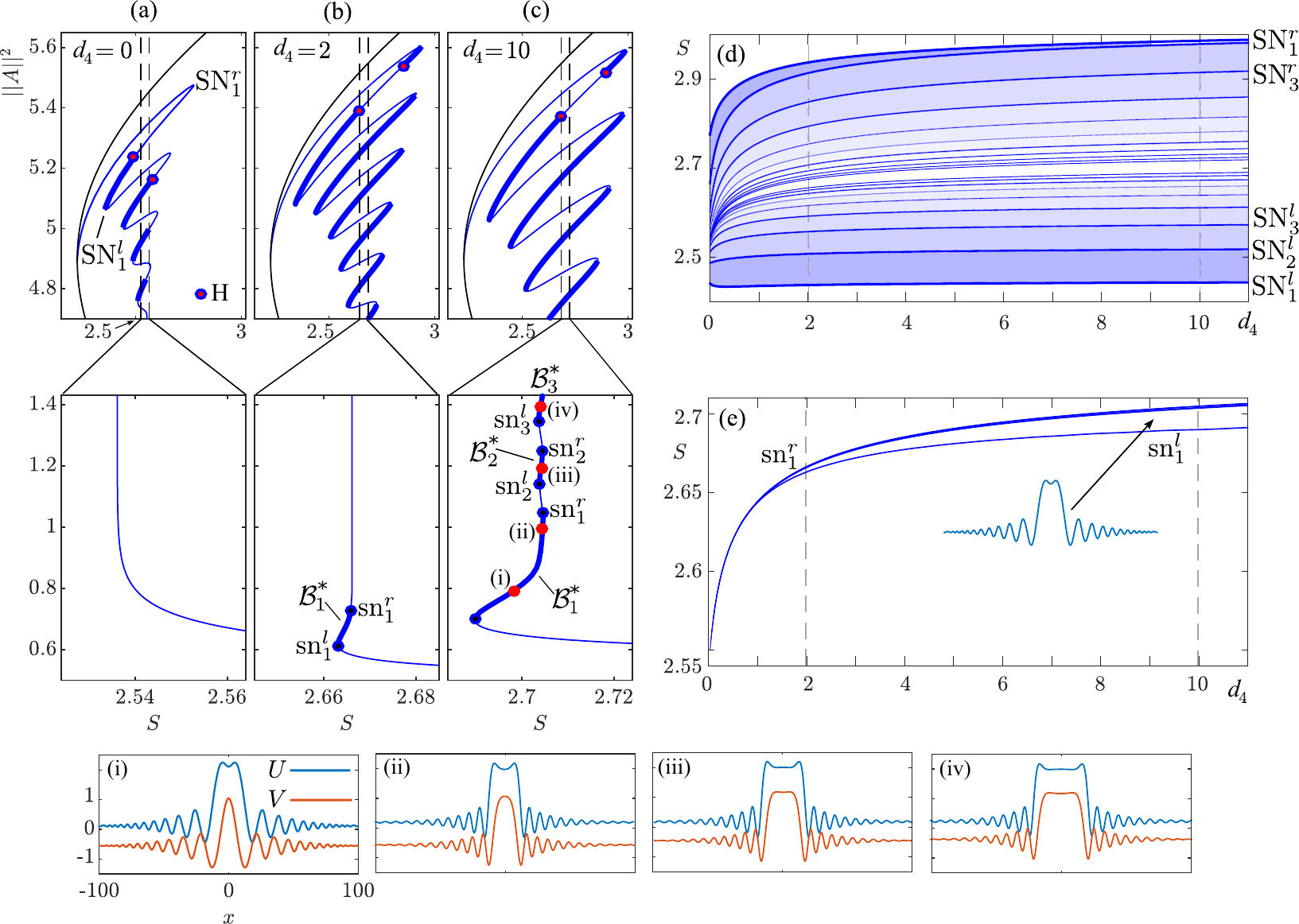}
\caption{Modification of the collapsed snaking for $\Delta=5$ when varying $d_4$: in (a) $d_4=0$, in (b) $d_4=2$, and $d_4=10$ in (c). Examples of bright LSs of different widths, corresponding to the diagram shown in (c), are depicted in panels (i)-(iv).
Panel (d) shows the 
$(\Delta,d_4)$-parameter phase diagram for dark LSs. Panel (e) shows the phase diagram associated with bright LSs. The vertical dashed lines correspond to the diagrams shown in panels (a)-(c).  }
\label{Dia_Beta_Delta}
\end{figure*}
Figure~\ref{breather2}(a) shows a close-up view of F$_2$ where these transitions are illustrated in more detail. The vertical dashed lines in Fig.~\ref{breather2}(a) correspond to the time series and frequency spectra shown below [see Figs.~\ref{breather2}(i)-(viii)], which represent the evolution of the breather intensity norm $||A||^2$ and its Fourier transform $||A||_f^2$, respectively.  From the modification of the spectra, one can clearly understand the transition between the different temporal dynamical regimes. Such modification is depicted in the color-map shown in Fig.~\ref{breather2}(b) all along F$_2$. 

At (i) the breather is characterized by a single frequency and their harmonics. At $b$, a new frequency arises [see Fig.~\ref{breather2}(b)], and the breather undergoes a beating phenomenon between the former and latter frequency.  In Fig.~\ref{breather2} (ii), the new frequency is marked using a red arrow and its difference with the former one with $\delta$. Because of this beating, the temporal series (see top panel) shows modulation in the amplitude, characterized by the beat period $2\pi/\delta$. At (iii), the $\delta$ is larger, resulting in a smaller modulation on the temporal series. Increasing $S$ further, the beating becomes more chaotic leading to the dynamical behavior shown in Fig.~\ref{breather2} (iii). Eventually this states dies out, leading to a single frequency breather associated with the time series shown in Fig.~\ref{breather2} (iv). Increasing more $S$, a sequence of PD bifurcation occur leading to the breather dynamics depicted in Fig.~\ref{breather2} (v)-(vii).
\section{Implications of fourth-order dispersion on the bifurcation structure and stability}\label{sec:4}
Previously, we have performed bifurcation analysis for a fixed value of $d_4$, namely $d_4=1$. Here we explore the effects of modifying $d_4$ on the stability of LSs, and on the their bifurcation structure. We will see that large values of $d_4$ may also lead to the emergence of bright LSs, similarly to the scenario described in Ref.~\cite{parra-rivas_coexistence_2017} for TOD.

Figure~\ref{Dia_Beta_Delta}
shows the modification of the collapse snaking diagram for $\Delta=5$ with increasing $d_4$: $d_4=0$ in Fig.~\ref{Dia_Beta_Delta}(a), $d_4=2$ in Fig.~\ref{Dia_Beta_Delta}(b), and $d_4=10$ in Fig.~\ref{Dia_Beta_Delta}(c). Regarding the top panels, we can see how by increasing $d_4$, the branches $\mathcal{B}_i$ become wider, leading to a wider existence region for the different dark LSs. Furthermore, $d_4$ has an stabilizing effect on the breather states emerging from $\mathcal{B}_1$ and $\mathcal{B}_2$. This phenomenon can be easily observed comparing Fig.~\ref{Dia_Beta_Delta}(a) and Fig.~\ref{Dia_Beta_Delta}(b). For $d_4=2$, $\mathcal{B}_2$ has stabilized completely while $\mathcal{B}_1$ partially. Both stabilizations occur due to the movement of the Hopf bifurcations along the branches.  

\begin{figure*}[!t]
\centering
\includegraphics[scale = 1]{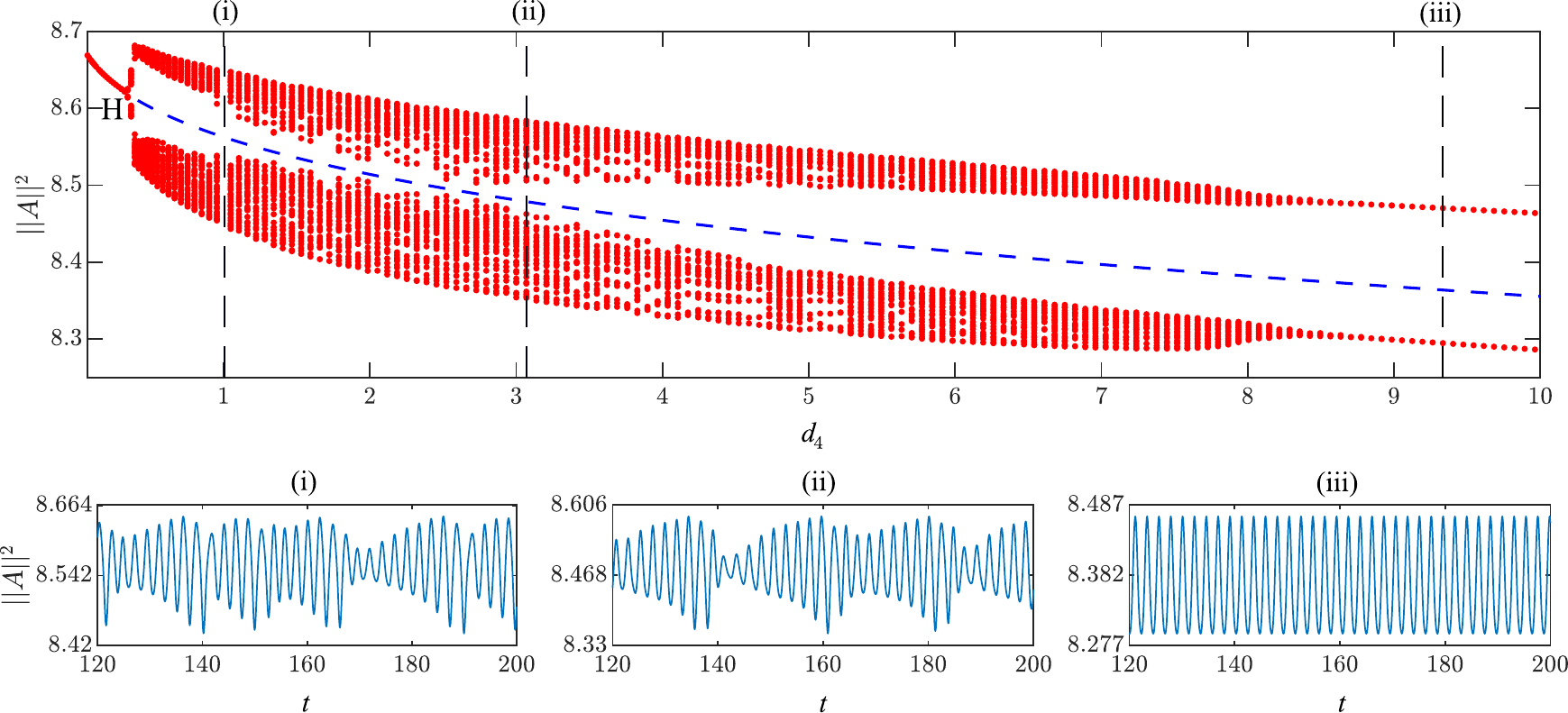}
\caption{Modification of the temporal dynamics associated with a single peak dark LSs when varying $d_4$. The plot shows the variation of the local maxima and minima of $||A||^2$  while passing from different dynamical regimes (red dots) and the unstable branch associated with the LS (dashed blue). The time series shown in panels (i)-(iii) correspond to the sections depicted through vertical dashed lines in the panel above. Here, $(\Delta,S)=(8,3.97)$. }
\label{beta4_Spatiotemporal_dynamics}
\end{figure*}
Performing a two parameter continuation of the different saddle-node bifurcations SN$_i^{l,r}$ in Fig.~\ref{Dia_Beta_Delta}(a)-(c)[top], we are able to compute their modification in parameters $S$ and $d_4$. The results are shown in the $(d_4,S)$-phase diagram shown in Fig.~\ref{Dia_Beta_Delta}(d). 
The vertical dashed lines correspond to the diagrams plotted in Fig.~\ref{Dia_Beta_Delta}(a)-(c)[top]. Decreasing, $d_4$,
the pair SN$_i^{l,r}$ (for $i$ fixed) comes closer, reducing the extension of $\mathcal{B}_i$ until reaching $d_4=0$. For $d_4< 0.5$ the shrinking process intensifies. In contrast, the pairs SN$_i^{l,r}$ separate softly with increasing $d_4$, and the separation seems to saturate for large values of $d_4$. The different blue tonalities between the SN$_i^{l,r}$ lines correspond to regions where dark LSs with different dips exist.
A similar tendency has also been observed for bright LSs in the presence of TOD \cite{parra-rivas_third-order_2014}, where in contrast, the saturation occurs when SN$_i^{l,r}$ approach each other with increasing $d_3$. 

Figures~\ref{Dia_Beta_Delta}(a)-(c) also show the modification, with increasing $d_4$, of the bottom part of the collapsed snaking around $S_M$. For $d_4=0$, this curve follows a straight line on top of $S_M$, which for $||A||^2\approx0.9$ turns to the right and eventually connects with MI at $A_h^b$ \cite{parra-rivas_dark_2016}. Increasing $d_4$ [see Fig.~\ref{Dia_Beta_Delta}(b) for $d_4=2$], the diagram reaches lower values of $||A||^2$, and undergoes a pair of saddle-node bifurcations sn$_1^{l,r}$ before reaching MI. Between these folds, a new branch of stable birght LS solutions ($\mathcal{B}^*_1$) appears. Increasing $\beta_4$ further [see Fig.~\ref{Dia_Beta_Delta}(c) for $d_4=10$], the extension of $\mathcal{B}^*_1$ increases. Two example of bright LSs on this branch are shown in Figs.~\ref{Dia_Beta_Delta}(i),(ii). Following up this diagram, new stable branches appear in-between sn$_2^{l,r}$ and sn$_3^{l,r}$. We label these branches $\mathcal{B}^*_2$ and $\mathcal{B}^*_3$, respectively. With increasing $||A||^2$, the bright LSs broaden as depicted in Figs.~\ref{Dia_Beta_Delta}(iii),(iv). The region of existence of these bright states is illustrated in the $(S,d_4)$-phase diagram shown in Fig.~\ref{Dia_Beta_Delta}(e). The saddle-node bifurcations sn$_i^{l,r}$ converge rapidly to $S_M$ for $i>1$. These states emerge due to the modification of the spatial eigenvalues, which now, allow the front locking around $A_h^b$ and $A_h^t$.

Finally, to complete our study, we analyse the effect of $d_4$ on the spatiotemporal dynamics of the system. To do so we fix $(\Delta,S)=(8,3.97)$, and scan the variation of the extrema of the dark LSs norm with $d_4$, as illustrated in Fig.~\ref{beta4_Spatiotemporal_dynamics}. For $d_4=0$, dark LSs are static, and the diagram just shows a single branch.  Increasing $d_4$ a bit further, the LSs encounter a Hopf bifurcation H, where they destabilize in favor of a single-period oscillation, which in a very narrow $d_4$-interval leads to more complex temporal dynamics characterized by the time series shown in Figs.~\ref{beta4_Spatiotemporal_dynamics}(i),(ii). Eventually, around $d_4\approx8.3$, the complex dynamics disappear, yielding a single period breather again [see temporal series in Fig.~\ref{beta4_Spatiotemporal_dynamics}(iii)].



\section{Discussions and Conclusions}\label{sec:5}
In this paper we have studied the formation, bifurcation structure and dynamical instabilities of dissipative Kerr LSs in the presence of FOD. Here we focus on the case where $d_2=1$ and $d_4>0$, a configuration which has not been studied in previous works. The linear stability analysis of the HSSs in this configuration shows the presence of bistability between two coexisting HSS states, namely $A_h^t$ and $A_h^b$ (see Sec.~\ref{sec:1}). We show that uniform wave fronts connecting the previous HSS states can lock, yielding the formation of LSs. For that, a necessary condition is the presence of oscillatory tails around either $A_h^t$ or $A_h^b$, which is determined by the spatial eigenvalues of the system. For $d_4=1$, our findings show that this locking is possible only around $A_h^b$, leading to the formation of dark LSs (see Sec.~\ref{sec:2.1}).

In bifurcation terms, these states undergo collapse homoclinic snaking whose extension and stability change with $\Delta$ (see Secs.~\ref{sec:2.2}, \ref{sec:2.3} and \ref{sec:3}). Collapsed homoclinic snaking is generic for systems exhibiting uniform-front locking, and has been reported in a variety of pattern forming systems including nonlinear quadratic optical resonators \cite{parra-rivas_localized_2019,arabi_localized_2020,parra-rivas_dissipative_2022,parra-rivas_dark_2021}, mode-locked vertical
external-cavity surface-emitting lasers (VCSEL)
\cite{schelte_tunable_2019,seidel_normal_2022}, semiconductor micro-resonators with strong
time-delayed feedback \cite{koch_temporal_2023},  and  reaction-diffusion systems \cite{zelnik_implications_2018,parra-rivas_formation_2020,al_saadi_unified_2021,ALSAADI2022,saadi_transitions_2022}, to only cite a few.

For $d_4=1$, the scenario presented here is morphologically very similar to the one found when studying Kerr cavities with only normal SOD (i.e., $d_2=1$) \cite{parra-rivas_dark_2016,parra-rivas_origin_2016}, despite of the extension of the LS existence regions and the onset of temporal instabilities. This result shows that dark states are robust against high-order dispersive effects.

This situation changes when increasing $d_4$ (Sec.~\ref{sec:4}). Regarding dark LSs, their region of existence initially increases, although their extension soon saturates for values of $d_4\approx3$. The most interesting phenomenon is that increasing $d_4$, bright LS solutions emerge, due to the stabilizing effect of this high-order dispersion term, and their existence region broadens. 
(see Fig.~\ref{Dia_Beta_Delta} in Sec.~\ref{sec:4}). We find that $d_4$ is also able to tune the emergence and type of oscillatory states appearing in the system, becoming a very relevant parameter to control the temporal dynamics.   

The capacity of $d_4$ for stabilizing LSs was discussed in Ref.~\cite{tlidi_high-order_2010} for a different regime of operation, where dark localized patterns, and their associated homoclinic snaking, were stabilized. The interaction of LSs, and formation of soliton molecules, in the presence of FOD effects have been also analyzed in Kerr cavities \cite{parra-rivas_interaction_2017,vladimirov_dissipative_2021}. In this context, FOD increases considerably the number of allowed locking distances between LSs, and therefore the variety of molecules. The stabilization of bright Kerr LSs in the context of uniform bistable regimes with collapsed snaking has been also investigated in the presence of TOD \cite{parra-rivas_coexistence_2017} and stimulated Raman scattering \cite{parra-rivas_influence_2021}. 

In summary, dark LSs formation is robust in the presence of FOD, which furthermore is able to stabilize, bright states. The advances in dispersion engineering make possible the management of different dispersion terms and therefore the access to dispersion regimes previously unattainable \cite{runge_pure-quartic_2020}. This capacity makes our work relevant not only from a theoretical point of view, but also from a experimental perspective, opening the possibility to observe these type of states and dynamics in real Kerr cavities.

 \section*{Acknowledgements}
 PPR acknowledges support from the European Union’s Horizon 2020 research and innovation programme
under the Marie Sklodowska-Curie grant agreement no. 101023717. FL acknowledges support from the European Research Council (grant agreement 757800), HORIZON EUROPE European Research Council (57800), and Fonds de la Recherche Scientifique-FNRS.

\bibliographystyle{ieeetr}
\bibliography{LSs_SHG}

\end{document}